\documentclass[
prd
,byrevtex
,preprint
,amsfonts,amsmath,amssymb
]{revtex4}
\usepackage{bm}
\usepackage[dvipdfmx]{graphicx}
\usepackage{ascmac}
\usepackage{amsthm}
\usepackage{latexsym}
\usepackage{physics}
\usepackage{color,float}
\theoremstyle{definition}
\newtheorem{theorem}{Theorem}[]
\newtheorem{corollary}{Corollary}[]
\newtheorem{lemma}{Lemma}[]

\begin{document}
\title{
\scalebox{0.9}{
Path-entangling evolution and quantum gravitational interaction}
}
\author{Akira Matsumura}
\email{matsumura.akira@phys.kyushu-u.ac.jp}
\affiliation{Department of Physics, Kyushu University, Fukuoka, 819-0395, Japan}

\begin{abstract}
We explore a general feature of the interaction mediated by the gravitational fields of spatially superposed masses. 
For this purpose, based on quantum information theory, we characterize the evolution of two particles each in a superposition state of paths. 
The evolution is assumed to be given by a completely positive trace-preserving (CPTP) map. 
We further assume that the probability of particle being on each path is unchanged during the evolution. 
This property is called population-preserving. 
We examine when a population-preserving CPTP map can create entanglement in terms of separable operations, which form a large class of local operations and classical communication (LOCC). 
In general, entanglement is not always generated by inseparable or non-LOCC operations, and one can consider a model of gravity described by an inseparable operation which does not create entanglement. 
However, we find that a population-preserving CPTP map is inseparable if and only if it can create entanglement.
This means that the above model of gravity is incompatible with the possible evolution of spatially superposed masses.
\end{abstract}
\maketitle

\tableofcontents
\section{Introduction}

Exploring the quantum nature of gravity is a fundamental task for modern physics. 
Recently, much of the theoretical and experimental efforts has been made to test the quantum superposition of the gravitational field caused by a spatially superposed mass. This phenomenon may be interpreted as the quantum superposition of a spacetime curvature.
To capture such an interesting phenomenon, gravity-induced entanglement has attracted attentions.

The gravity-induced entanglement is expected to be tested in table-top experiments. 
In Refs.\cite{Bose2017,Marletto2017}, it was shown that the gravitational interaction of two masses generates the entanglement between them when each mass travels through an individual matter-wave interferomenter. 
The works \cite{Bose2017,Marletto2017} has been stimulated advanced research and other proposals using matter-wave interferometers \cite{Nguyen2020, Chevalier2020, vandeKamp2020, Toros2021, Miki2021a}, levitated nanoparticles or mechanical oscillators \cite{Qvafort2020, Krisnanda2020}, optomechanical systems \cite{Balushi2018, Miao2020, Matsumura2020, Miki2021b} and their hybrid model \cite{Carney2021a, Carney2022, Pedernales2021, Matsumura2021b}.
Also, the field theoretical aspects of gravity-induced entanglement have been discussed in \cite{Belenchia2018, Marshman2020, Matsumura2021a, Carney2021b, Danielson2021, Bose2022, Christodoulou2022}.  

The theoretical importance of gravity-induced entanglement mainly relies on quantum information theory. 
To present it explicitly, the paradigm of local operations and classical communication (LOCC) plays a crucial role \cite{Nielsen2002, Horodecki2009}. 
An LOCC operation is a physical process in two distant systems, which describes the local evolution of each system and the causal transfer of classical information between them.
In order to capture general properties of LOCC operations, one often introduces the notion of separable operations \cite{Rains1997, Vedral1998} including all LOCC operations as a proper subset \cite{Chitambar2014}. 
It is known that any separable (or LOCC) operation cannot generate entanglement in non-entangled (separable) states \cite{Horodecki2009}.
Gravity-induced entanglement is an evidence that a gravitational interaction is described by an inseparable or non-LOCC operation.
In terms of LOCC, if the interaction is local, then the gravity-induced entanglement is incompatible with classical communication. 
This means that the gravitational field carries quantum information.

As mentioned above, entanglement tells us the description of gravity in quantum information theory, that is, whether gravity is described by an inseparable (or a non-LOCC) operation.
However, the relation between entanglement and separable operations is not simple: there is an inseparable operation not to generate entanglement \cite{Cirac2001, Harrow2003}. 
In principle, one can consider a model of the gravitational interaction represented by an inseparable operation without entanglement creation. 

In this paper, to discuss such a possibility, we examine a general feature of the quantum interaction between two particles. 
The tools in quantum information theory make the analysis independent of the details of interaction.
As in Ref.\cite{Bose2017,Marletto2017}, we consider the setting that each particle is in a superposition of paths.
The evolution of the particles is assumed to be given by a completely positive trace-preserving (CPTP) map \cite{Nielsen2002, Horodecki2009}, which describes a large class of physical dynamics.  
We further suppose that the probability of particle being on each path is preserved.
This property is called population-preserving in \cite{Carney2021a}. 
We can derive a representation theorem for a population-preserving CPTP map.
To characterize the entanglement evolution of the particles, we use the notion of a separable operation and a non-entangling operation \cite{Harrow2003}. 
The latter is defined as a CPTP map which transforms every separable state to another separable state (that is, entanglement does not occur). 
It is known that those operations are not equivalent to each other \cite{Harrow2003}. 
However, we show the equivalence between inseparable operation and entangling operation if they are population-preserving CPTP.
Within the above framework and assumptions, this means that gravity described by an inseparable operation without entanglement creation is incompatible with the evolution of spatially superposed masses.

The structure of this paper is as follows. 
In Sec. \ref{sec:dynamics}, the population-preserving dynamics of superposed particles is presented. 
In Sec. \ref{sec:Demo}, we demonstrate the entanglement generation by the gravitational potential between two massive particles. 
In Sec. \ref{sec:separable}, we obtain several key properties of a population-preserving CPTP map and show the equivalence between inseparable and entangling operations.  
Sec. \ref{sec:Conclusion} presents the conclusion of this paper. 
We use the unit
$\hbar=1$ in this paper. 


\section{Population-preserving dynamics of superposed particles}
\label{sec:dynamics}

\begin{figure}[H]
  \centering
  \includegraphics[width=0.50\linewidth]{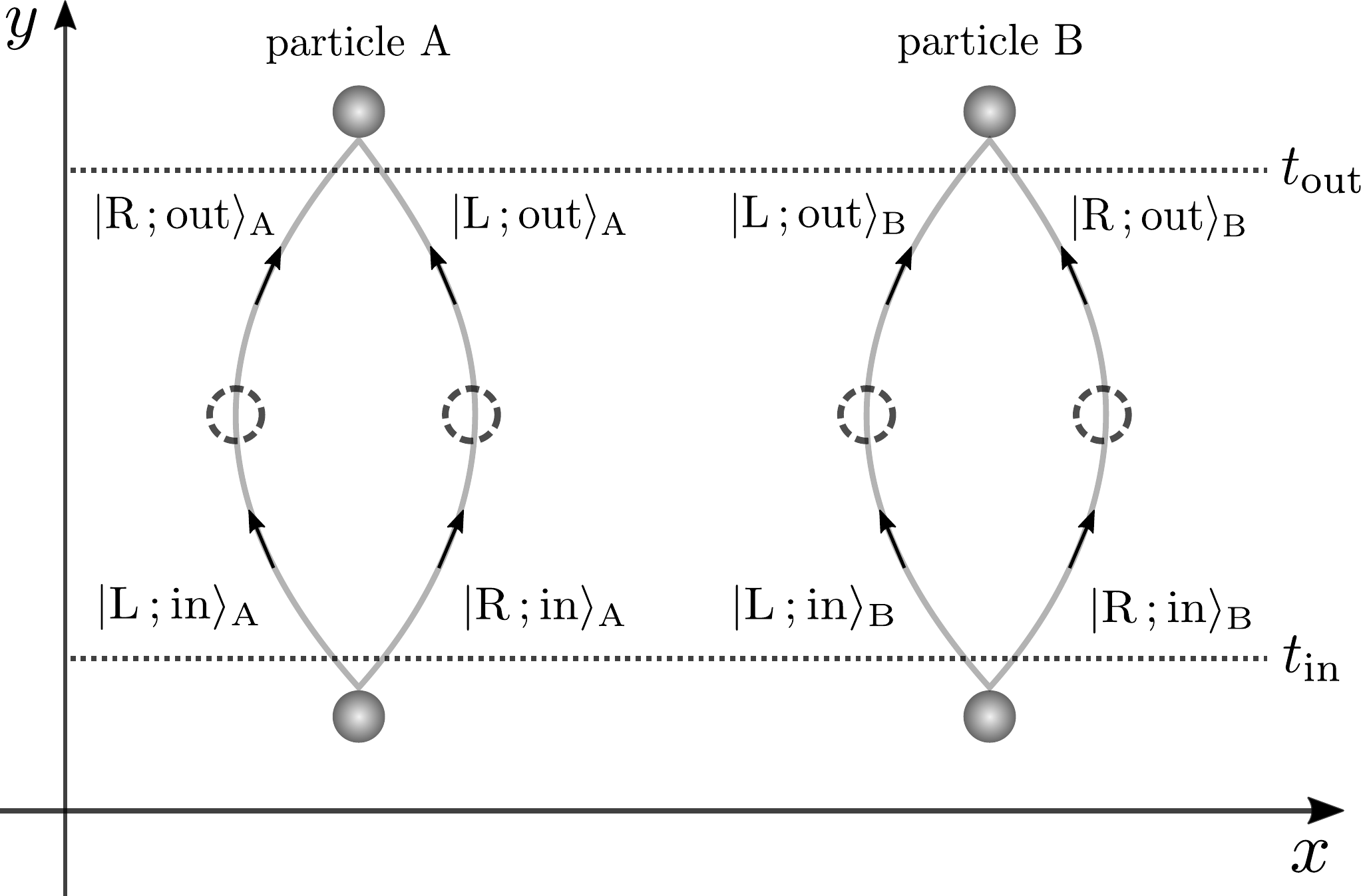}
  \caption{An example of the configuration of two particles A and B.}
  \label{fig:conf1}
\end{figure}

In this section, to show the population-preserving property, we consider the dynamics of two particles A and B described by the Schr\"{o}dinger equation with the Hamiltonian 
\begin{equation}
\hat{H}=\hat{H}_\text{A}+\hat{H}_\text{B}+\hat{V},
\label{eq:H}
\end{equation}
where 
$\hat{H}_\text{A}$ and 
$\hat{H}_\text{B}$ are each Hamiltonian of the particles A and B, and 
$\hat{V}$ is the interaction between them.
Each particle is assumed to be superposed on two paths (see Fig.\ref{fig:conf1}). 
Here, we also assume that each path is determined by the Hamiltonians 
$\hat{H}_\text{A}$ and 
$\hat{H}_\text{B}$ and is 
almost undisturbed by the interaction 
$\hat{V}$.
Each particle is initially in the following superposed state, 
\begin{equation}
|\Psi_\text{in} \rangle = \sum_{a,b=\text{L},\text{R}}  \psi_a |a\,;\text{in} \rangle_\text{A} \, \phi_b |b\,;\text{in} \rangle_\text{B},
\label{eq:Psii}
\end{equation}
where 
$|a\,;\text{in} \rangle_\text{A} \, (|b\,;\text{in} \rangle_\text{B})$ satisfying the orthonormal condition 
${}_\text{A} \langle a';\text{in} |a\,;\text{in} \rangle_\text{A} \approx \delta_{a'a}\, (\, {}_\text{B} \langle b';\text{in}|b\,;\text{in}\rangle_\text{B} \approx \delta_{b'b}\,)$
represents the state of the particle A (B) with the localized wave packet whose peak is at the position $a \, (b)$ at an initial time 
$t_\text{in}$, and 
$\psi_a\, (\phi_b)$ gives the probability
$|\psi_a|^2 \, (|\phi_b|^2)$ which the particle A (B) is in $|a\,;\text{in} \rangle_\text{A} \, (\, |b\,;\text{in} \rangle_\text{B}\, )$. 
The evolved state of the particles A and B is 
\begin{align}
|\Psi_\text{out} \rangle 
&= e^{-i\hat{H}(t_\text{out}-t_\text{in})}|\Psi_\text{in}\rangle
\nonumber 
\\
&=
e^{-i(\hat{H}_\text{A}+\hat{H}_\text{B})(t_\text{out}-t_\text{in})}\text{T}
\exp \Big[-i\int^{t_\text{out}}_{t_\text{in}} dt \hat{V}_\text{I} (t) \Big] |\Psi_\text{in} \rangle 
\nonumber 
\\
&=
e^{-i(\hat{H}_\text{A}+\hat{H}_\text{B})(t_\text{out}-t_\text{in})}\text{T}
\exp \Big[-i\int^{t_\text{out}}_{t_\text{in}} dt \hat{V}_\text{I} (t) \Big] \sum_{a,b=\text{L},\text{R}}\psi_a |a\,;\text{in} \rangle_\text{A} \, \phi_b |b\,;\text{in} \rangle_\text{B}
\nonumber 
\\
&\approx
e^{-i(\hat{H}_\text{A}+\hat{H}_\text{B})(t_\text{out}-t_\text{in})} \sum_{a,b=\text{L},\text{R}}e^{i\Phi_{ab}}\, \psi_a |a\,;\text{in} \rangle_\text{A} \, \phi_b |b\,;\text{in} \rangle_\text{B}
\nonumber 
\\
&=
\sum_{a,b=\text{L},\text{R}}e^{i\Phi_{ab}}\, \psi_a |a\,;\text{out}\rangle_\text{A} \, \phi_b |b\,;\text{out} \rangle_\text{B}, 
\label{eq:Sol}
\end{align}
where 
$\hat{V}_\text{I}(t)$ is the interaction operator in the interaction picture with respect to 
$\hat{H}_\text{A}+\hat{H}_\text{B}$, and the out states
\begin{align}
|a\,;\text{out}\rangle_\text{A} = e^{-i\hat{H}_\text{A}(t_\text{out}-t_\text{in})}|a\,;\text{in} \rangle_\text{A}, \quad |b\,;\text{out} \rangle_\text{B} = e^{-i\hat{H}_\text{B}(t_\text{out}-t_\text{in})}|b\,;\text{in} \rangle_\text{B}
\label{eq:out}
\end{align}
describe the wave packets of the particles A and B with the peaks at the positions 
$a$ and 
$b$ at the time 
$t_\text{out}$, respectively. 
In the fourth line of \eqref{eq:Sol}, since the motion of each particle is unperturbed, the interaction potential 
$\hat{V}$ is evaluated along each path, which leads to the phase factors 
$e^{i\Phi_{ab}}$ with
\begin{equation}
\Phi_{ab}=-\int^{t_\text{out}}_{t_\text{in}} dt V_{ab} (t).
\label{eq:PhiPQ}
\end{equation}
The kinetic and other potential terms in the local Hamiltonians 
$\hat{H}_\text{A}$ and 
$\hat{H}_\text{B}$ may give another accumulated phases, which are included in the out states 
$|a\,;\text{out}\rangle_\text{A}$ and 
$|b\,;\text{out} \rangle_\text{B}$. 
Since such phases are given by a local unitary evolution, they do not affect the entanglement between the particles.
However, the fluctuation of the phases may cause dephasing or decoherence effects. 
The detailed analysis of such effects was performed in \cite{Toros2021}. 

We consider two particles with masses $m_\text{A}$ and
$m_\text{B}$. 
The gravitational potential between the massive particles is 
\begin{equation}
\hat{V}
= -\frac{Gm_\text{A}m_\text{B}}{|\hat{\bm{x}}_\text{A} - \hat{\bm{x}}_\text{B}|},
\label{eq:1/r}
\end{equation}
where 
$\hat{\bm{x}}_\text{A}$ and 
$\hat{\bm{x}}_\text{B}$ are the position operators of each particle. 
The potential energy 
$V_{ab}(t)$ in the accumulated phase \eqref{eq:PhiPQ} is given as
\begin{equation}
V_{ab}(t)= -\frac{Gm_\text{A}m_\text{B}}{|\bm{x}^a_\text{A}(t) - \bm{x}^b_\text{B}(t)|}.
\label{eq:VPQ}
\end{equation}
Here, 
$\bm{x}^a_\text{A}(t)$ and 
$\bm{x}^b_\text{B}(t)$ describe the trajectories of the massive particles A and B which initially sit at the positions 
$a$ and $b$, respectively. 
One notes that the gravitational interaction is instantaneous and the locality (causality) of theory seems not to satisfy. 
We should consider that the gravitational interaction is just given in the non-relativistic limit and the locality holds in principle. 
This concern was recently discussed in \cite{Bose2022, Christodoulou2022}. 
In Sec. \ref{sec:Demo}, the gravity-induced entanglement between A and B will be demonstrated. 

From the evolution given in Eq.\eqref{eq:Sol}, we find the equality
\begin{equation}
 \langle \Psi_\text{out}|\hat{\Pi}^\text{out}_a \otimes  \hat{\Pi}^\text{out}_b |\Psi_\text{out} \rangle
 =\langle \Psi_\text{in}| \hat{\Pi}^\text{in}_a \otimes  \hat{\Pi}^\text{in}_b |\Psi_\text{in} \rangle,
 \label{eq:PP}
\end{equation}
where 
$\hat{\Pi}^\text{in}_a=|a\,;\text{in} \rangle_\text{A} \langle a\,;\text{in}|$,
$\hat{\Pi}^\text{out}_a=|a\,;\text{out}\rangle_\text{A} \langle a\,;\text{out}|$, 
$\hat{\Pi}^\text{in}_b=|b\,;\text{in}\rangle_\text{B} \langle b\,;\text{in}|$
and 
$\hat{\Pi}^\text{out}_b=|b\,;\text{out}\rangle_\text{B} \langle b\,;\text{out}|$. 
Namely, the probability which we find the particles on each path is preserved.
This property is called population-preserving in Ref.\cite{Carney2021a}. 
In Sec. \ref{sec:separable}, we will introduce a completely positive trace-preserving map, a separable operation and a non-entangling operation. 
Imposing the population-preserving property, we will examine how the map and the operations are represented. 

\section{Demonstration of gravity-induced entanglement}
\label{sec:Demo}

In this section, the basics of entanglement are introduced, and the entanglement due to gravitational interaction is demonstrated. 
To quantify entanglement, we introduce the positive partial
transpose criterion \cite{Peres1996, Horodecki1996} and the negativity \cite{Vidal2002} as follows. 
For a given density operator 
$\rho$ of a bipartite system AB, we define the partial transposition
$\rho^{\text{T}_\text{A}}$ with the components
\begin{equation}
{}_\text{A} \langle a |  {}_\text{B} \langle b| \rho^{\text{T}_\text{A}} |a' \rangle_\text{A} |b' \rangle_\text{B}= {}_\text{A} \langle a' |  {}_\text{B} \langle b| \rho |a \rangle_\text{A}  |b' \rangle_\text{B}
\end{equation}
for a basis 
$\{ |a \rangle_\text{A} |b \rangle_\text{B} \}_{a,b}$ of the Hilbert space $\mathcal{H}_\text{A} \otimes \mathcal{H}_\text{B}$ of the bipartite system AB.  
One can show that if the density operator 
$\rho$ is separable (non-entangled) \cite{Werner1989}, that is, 
\begin{equation}
\rho=\sum_{k} p_k \rho_k \otimes \sigma_k,
\label{eq:sep}
\end{equation}
where 
$p_k$ is a probability, and 
$\rho_k$ and 
$\sigma_k$ are the density operators of each subsystem, then the partial transposition 
$\rho^{\text{T}_\text{A}}$ has only non-negative eigenvalues. 
Hence, if the partial transposition 
$\rho^{\text{T}_\text{A}}$ has a negative eigenvalue then the given density operator $\rho$ is non-separable (entangled). 
This is called the positive partial transpose (PPT) criterion \cite{Peres1996, Horodecki1996}. 
In particular, the PPT criterion becomes the necessary and sufficient condition for a two-qubit system 
($\mathbb{C}^2 \otimes \mathbb{C}^2$) and a qubit-qutrit system 
($\mathbb{C}^2 \otimes \mathbb{C}^3$) \cite{Horodecki1996}. 
The negativity 
\begin{equation}
\mathcal{N}=\sum_{\lambda_i<0} |\lambda_i|
\end{equation}
with the eigenvalues 
$\lambda_i$ of the partial transposition $\rho^{\text{T}_\text{A}}$ is an entanglement measure (quantifier) \cite{Vidal2002}. 
The PPT criterion means that a density operator 
$\rho$ with nonzero negativity is entangled. 

\begin{figure}[H]
  \centering
  \includegraphics[width=0.50\linewidth]{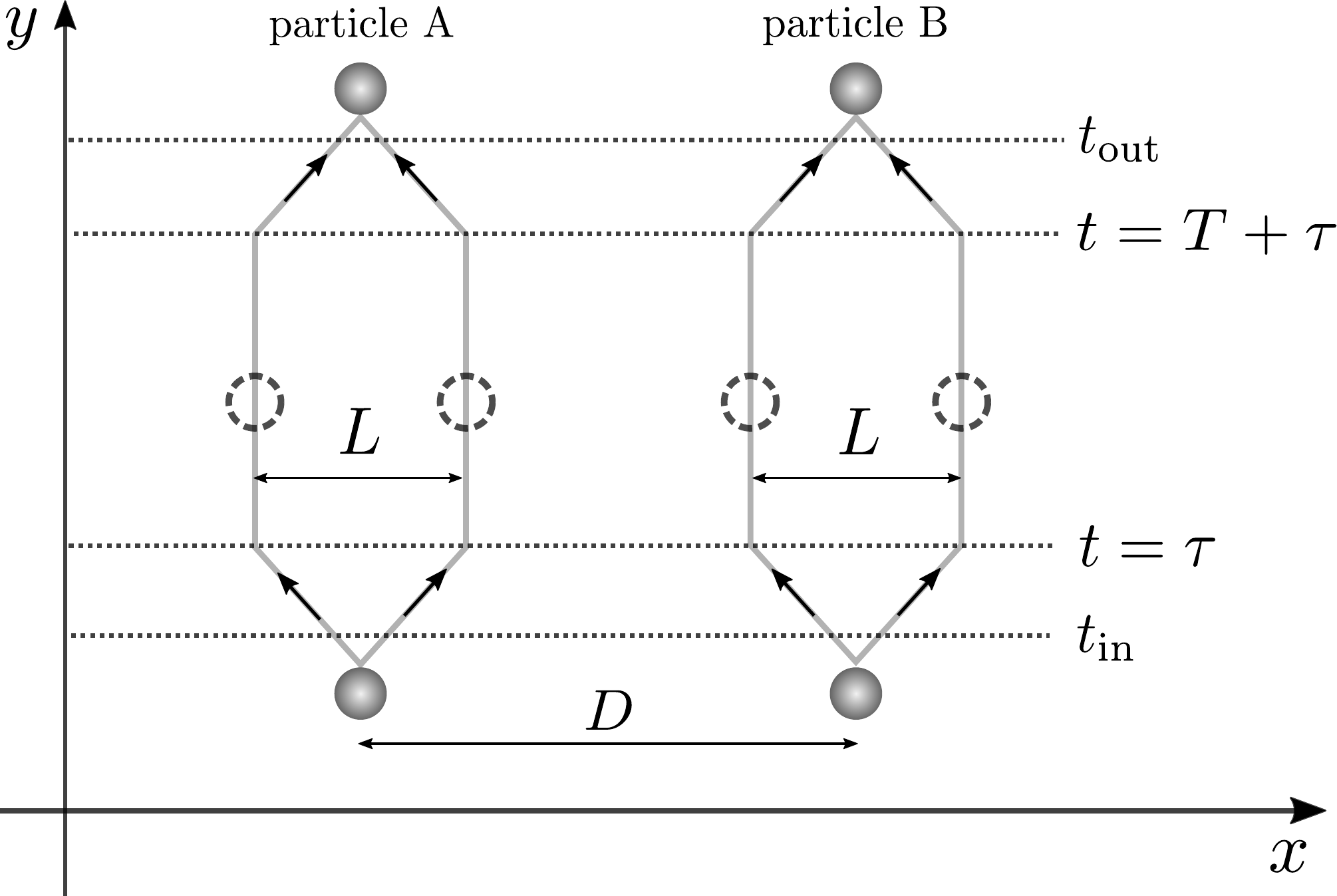}
  \caption{The concrete configuration to demonstrate the entanglement behavior of two particles. 
  $L$ is the typical length scale of superposition of each particle, and 
  $D$ is the distance between the two particles at the initial time $t_\text{in}$ }
  \label{fig:conf2}
\end{figure}

\begin{figure}[H]
  \centering
  \includegraphics[width=0.80\linewidth]{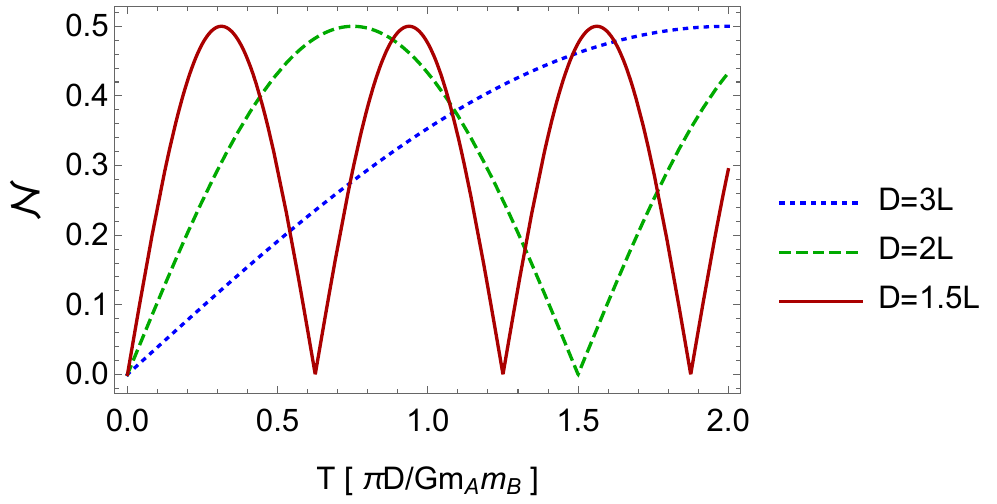}
  \caption{The behavior of the negativity between the two massive particles.
  The vertical axis represents the negativity, which is a dimensionless quantity. 
  The horizontal axis describes the time 
  $T$ in the unit 
  $\pi D /Gm_\text{A}m_\text{B}$. 
  }
  \label{fig:ngtvty}
\end{figure}

Let us demonstrate the entanglement generation due to the gravitational potential \eqref{eq:1/r} assuming the concrete trajectories as 
\begin{equation}
\bm{x}^a_\text{A}(t)=[x^a(t), v_y t,0]^\text{T}, \quad \bm{x}^b_\text{B}(t)=[x^b(t)+D, v_y t,0]^\text{T},
\end{equation}
where 
$v_y>0$ is the velocity in the y-direction and 
\begin{equation}
x^a(t)=\epsilon_a
\left \{
\begin{array}{lll}
 vt & t_\text{in} \leq t \leq \tau \\
 v\tau & \tau \leq t \leq T+\tau \\ 
 v(T+\tau-t)+v\tau & T+\tau \leq t \leq t_\text{out}
\end{array}
\right.
\end{equation}
with 
$\epsilon_\text{L}=-\epsilon_\text{R}=-1$ and 
the velocity 
$v(>0)$ in the x-direction. 
The accumulated phase \eqref{eq:PhiPQ} is computed as
\begin{align}
\Phi_{ab}= \int^{t_\text{out}}_{t_\text{in}} dt \frac{ Gm_\text{A}m_\text{B}}{|\bm{x}^a_\text{A}(t) - \bm{x}^b_\text{B}(t)|}
\approx \int^{T+\tau}_{\tau} dt \frac{ Gm_\text{A}m_\text{B}}{|\bm{x}^a_\text{A}(t) - \bm{x}^b_\text{B}(t)|}
= 
\frac{Gm_\text{A}m_\text{B}T}{D+(\epsilon_b-\epsilon_a)L/2} ,
\label{eq:PhiPQ2}
\end{align}
where 
$L=2v\tau$. 
In the approximation, assuming a large 
$T$, we evaluated the phase given in the intermediate stage 
$\tau \leq t \leq T+\tau$. 
The configuration of the particle paths is shown in Fig.\ref{fig:conf2}.
Assuming the initial state \eqref{eq:Psii} with 
$\psi_a=\phi_b=1/\sqrt{2}$, we obtain the following negativity 
\begin{equation}
\mathcal{N}=\frac{1}{2}
 \Big|
 \sin
\Big[
\frac{\Phi_\text{RL}+\Phi_\text{LR}-\Phi_\text{LL}-\Phi_\text{RR}}{2}
\Big]\Big|
=\frac{1}{2}
 \Big|
 \sin
\Big[
\frac{Gm_\text{A}m_\text{B}T}{2} \Big( \frac{1}{D+L}+\frac{1}{D-L}-\frac{2}{D} \Big)
\Big]\Big|.
 \label{eq:expN}
\end{equation}
Fig.\ref{fig:ngtvty} shows the negativity as a function of 
$Gm_\text{A}m_\text{B}T/\pi D$ for a fixed 
$D$.
For a small 
$D$, the accumulated phases
frequently change, and the entanglement generation due to the interaction is promoted. 

\section{Physical operations on spatially superposed particles}
\label{sec:separable}

In this section, the main theorems for the evolution of two particles in a path superposition are devoted.
In the subsection \ref{subsec:CPTPA}, assuming the evolution is described by a completely positive trace-preserving and population-preserving map, we will obtain a representation theorem for the evolution. 
In the subsection \ref{subsec:sepB}, introducing the notion of separable operations, we will find the separability criterion on the evolution. 
In the subsection \ref{subsec:eqC}, it will be shown that inseparable operations on the particles are equivalent to entangling operations. 

\subsection{Characterization of completely positive trace-preserving maps}
\label{subsec:CPTPA}

We consider a map 
$\Phi$ from the set of density operators on a Hilbert space 
$\mathcal{H}$ to the set of density operators on another Hilbert space 
$\mathcal{K}$. 
The map 
$\Phi$ is assumed to satisfy the following properties:
$\text{Tr}\big[\Phi[\rho] \big]=\text{Tr}[\rho]$ (trace preserving), 
$\Phi[\lambda \rho + (1-\lambda)\sigma]
=\lambda \Phi[\rho]+(1-\lambda)\Phi[\sigma]$ (convex linear) and
$\Phi \otimes \mathcal{I}_\text{E}[\Omega] \geq 0$ (completely positive), where 
$\rho$ and 
$\sigma$ are density operators on 
the Hilbert space $\mathcal{H}$, 
$\lambda$ is a probability 
($0 \leq \lambda \leq 1$), 
$\mathcal{I}_\text{E}$ is the identity map on an extra system E with a Hilbert space 
$\mathcal{H}_\text{E}$, and 
$\Omega$ is an arbitrary density operator on the Hilbert space
$\mathcal{H} \otimes \mathcal{H}_\text{E}$.
The map 
$\Phi$ is called a completely positive trace-preserving (CPTP) map \cite{Nielsen2002,Horodecki2009}.
The CPTP map 
$\Phi$ can describe a unitary evolution 
($\Phi[\rho]=\hat{U} \rho \hat{U}^\dagger$) and the coupled dynamics with measurement apparatuses or environments.  
It is known that every CPTP map is represented as 
\begin{equation}
\Phi[\rho]=
 \sum_\ell \hat{K}_\ell \, \rho \, \hat{K}^\dagger_\ell, 
\label{eq:Kraus}
\end{equation}
where the operators
$\hat{K}_\ell : \mathcal{H} \rightarrow \mathcal{K}$ satisfying 
$\sum_\ell \hat{K}^\dagger_\ell \hat{K}_\ell=\hat{\mathbb{I}}_\mathcal{H}$ ( $\hat{\mathbb{I}}_\mathcal{H}$ is the indentity operator on 
$\mathcal{H}$)
are called Kraus operators. 
Conversely, the map defined by Eq.\eqref{eq:Kraus} is a CPTP map. 
The representation Eq.\eqref{eq:Kraus} of a CPTP map is called the operator-sum representation \cite{Nielsen2002}.

In the setting presented in Sec.\ref{sec:dynamics} (for example, see Fig.\ref{fig:conf1}), the population preserving holds, that is,
the probability of particles being on each path does not change. 
We assume that the evolution of the particles is given by a population-preserving CPTP map $\Phi$. 
The latter condition means that the following equality holds for any initial density operator 
$\rho_\text{in}$: 
\begin{equation}
 \text{Tr}\big[ \hat{\Pi}^\text{out}_a\otimes  \hat{\Pi}^\text{out}_b \rho_\text{out} \big]
 =\text{Tr}\big[ \hat{\Pi}^\text{in}_a \otimes  \hat{\Pi}^\text{in}_b \rho_\text{in} \big],
 \label{eq:PP2}
\end{equation}
where 
$\hat{\Pi}^\text{in}_a=|a\,;\text{in}\rangle_\text{A} \langle a\,;\text{in}|$, 
$\hat{\Pi}^\text{out}_a=|a\,;\text{out}\rangle_\text{A} \langle a\,;\text{out}|$,
$\hat{\Pi}^\text{in}_b=|b\,;\text{in}\rangle_\text{B} \langle b\,;\text{in}|$, 
$\hat{\Pi}^\text{out}_b=|b\,;\text{out}\rangle_\text{B} \langle b\,;\text{out}|$ and 
$\rho_\text{out}=\Phi[\rho_\text{in}]$.
In Sec.\ref{sec:dynamics}, each particle was assumed to be superposed in two paths. 
One can consider the setting of a multipath superposition, which is adopted to test gravity-induced entanglement \cite{Tilly2021}. 
Taking into account multiple paths of each particle, we can show the following theorem. 
\begin{theorem}
Let 
$\mathcal{H}^\text{in}_\text{A} \otimes \mathcal{H}^\text{in}_\text{B}$ and 
$\mathcal{H}^\text{out}_\text{A} \otimes \mathcal{H}^\text{out}_\text{B}$
be the Hilbert space with the complete bases 
$\{ |a\,;\text{in} \rangle_\text{A} |b\,;\text{in} \rangle_\text{B} \}_{a=1,2,\dots,d_\text{A},b=1,2,\dots,d_\text{B}}$ and 
$\{ |a\,;\text{out}\rangle_\text{A} |b\,;\text{out} \rangle_\text{B} \}_{a=1,2,\dots,d_\text{A},b=1,2,\dots,d_\text{B}}$, respectively. 
A map 
$\Phi :\rho_\text{in} \mapsto \rho_\text{out}=\Phi[\rho_\text{in}] $, where 
$\rho_\text{in}$ and 
$\rho_\text{out}$ are density operators each on  
$\mathcal{H}^\text{in}_\text{A} \otimes \mathcal{H}^\text{in}_\text{B}$ and
$\mathcal{H}^\text{out}_\text{A} \otimes \mathcal{H}^\text{out}_\text{B}$, 
is CPTP and population-preserving if and only if the map
$\Phi$ is represented by using  
$\hat{M}_a=|a\,;\text{out}\rangle_\text{A} \langle a\,;\text{in}|$ and  
$\hat{N}_b=|b\,;\text{out}\rangle_\text{B} \langle b\,;\text{in}|$ as 
\begin{equation}
\Phi[\rho_\text{in}]
 =\sum^{d_\text{A}}_{a,a'=1} 
 \sum^{d_\text{B}}_{b,b'=1} \mathcal{E}_{aba'b'} 
 \hat{M}_a \otimes  \hat{N}_b \, \rho_\text{in}  \, 
 \hat{M}^\dagger_{a'}  \otimes  \hat{N}^\dagger_{b'},
\label{eq:CPTPPP}
\end{equation}
where the coefficients 
$\mathcal{E}_{aba'b'} $ 
satisfy
$\mathcal{E}_{abab}=1$, and the square matrix 
$\mathcal{E}$ with the components $\mathcal{E}_{aba'b'}$ has only non-negative eigenvalues, whose characteristic equation with an eigenvector
$\bm{v}=[v_{11},v_{12},\dots,v_{1d_\text{B}}, v_{21}, v_{22}, \dots, v_{d_\text{A}d_\text{B}} ]^\text{T}$
and an eigenvalue
$\nu$ is 
\begin{equation}
(\mathcal{E}\bm{v})_{ab}
=\sum^{d_\text{A}}_{a'=1} 
 \sum^{d_\text{B}}_{b'=1} \mathcal{E}_{aba'b'} v_{a'b'} =\nu v_{ab}.
\end{equation}
The latter condition on the matrix 
$\mathcal{E}$ is equivalent to that the components
$\mathcal{E}_{aba'b'}$ satisfy
$\sum_{a,b,a',b'} Z_{ab}\mathcal{E}_{aba'b'}Z^*_{a'b'} \geq 0 $ for all complex numbers $Z_{ab}$. 
\begin{proof}

Let us first prove the sufficient condition. 
A population-preserving CPTP map
$\Phi$ satisfies 
\begin{align}
\text{Tr}\big[ \hat{\Pi}^\text{in}_a \otimes  \hat{\Pi}^\text{in}_b \rho_\text{in} \big] 
&=
\text{Tr}\big[ \hat{\Pi}^\text{out}_a\otimes  \hat{\Pi}^\text{out}_b \Phi[\rho_\text{in}] \big]
\nonumber 
\\
&=
\text{Tr}\big[ \hat{\Pi}^\text{out}_a\otimes  \hat{\Pi}^\text{out}_b \sum_\ell \hat{K}_\ell \, \rho_\text{in} \, \hat{K}^\dagger_\ell  \big]
\nonumber 
\\
&=
\text{Tr}
\Big[\Big(\sum_\ell
\hat{K}^\dagger_\ell \, \hat{\Pi}^\text{out}_a\otimes  \hat{\Pi}^\text{out}_b \,  \hat{K}_\ell\Big) \, \rho_\text{in}
\Big],
\end{align}
where 
$\hat{\Pi}^\text{in}_a=|a\,;\text{in}\rangle_\text{A} \langle a\,;\text{in}|$, 
$\hat{\Pi}^\text{out}_a=|a\,;\text{out}\rangle_\text{A} \langle a\,;\text{out}|$,
$\hat{\Pi}^\text{in}_b=|b\,;\text{in}\rangle_\text{B} \langle b\,;\text{in}|$, 
$\hat{\Pi}^\text{out}_b=|b\,;\text{out}\rangle_\text{B} \langle b\,;\text{out}|$, and 
the first equality holds by the population preserving, and in the second line the operator-sum representation \eqref{eq:Kraus} of 
$\Phi$ was used. 
The above equality holds for any initial states. 
Choosing the initial state as
$\rho_\text{in}=|a'\,;\text{in}\rangle_\text{A} \langle a'\,;\text{in}| \otimes |b';\text{in}\rangle_\text{B} \langle b';\text{in}|$, we get the following equation
\begin{equation}
\delta_{aa'} \delta_{bb'}=
\sum_\ell |{}_\text{A} \langle a\,;\text{out}| {}_\text{B} \langle b\,;\text{out}| \hat{K}_\ell |a'\,;\text{in} \rangle_\text{A} |b';\text{in} \rangle_\text{B}|^2, 
\end{equation}
where 
$\delta_{aa'}$ is the Kronecker's delta.
For 
$a\neq a'$ or 
$b\neq b'$, the above equation yields 
${}_\text{A} \langle a\,;\text{out}| {}_\text{B} \langle b\,;\text{out}| \hat{K}_\ell |a'\,;\text{in} \rangle_\text{A} |b';\text{in} \rangle_\text{B}=0$. 
Hence, the Kraus operator 
$\hat{K}_\ell$ has the form 
\begin{equation}
 \hat{K}_\ell=\sum^{d_\text{A}}_{a=1} 
 \sum^{d_\text{B}}_{b=1} k^{(\ell)}_{ab} \,\hat{M}_a \otimes  \hat{N}_b, 
 \label{eq:Kell}
\end{equation}
where 
$\hat{M}_a=|a\,;\text{out}\rangle_\text{A} \langle a\,;\text{in}|$,
$\hat{N}_b=|b\,;\text{out} \rangle_\text{B} \langle b\,;\text{in}|$ and a complex number 
$k^{(\ell)}_{ab}$. 
Then, the map
$\Phi$ is represented as
\begin{equation}
\Phi[\rho]
 =\sum_{\ell} \hat{K}_\ell \rho \hat{K}^\dagger_\ell
 =\sum^{d_\text{A}}_{a,a'=1} 
 \sum^{d_\text{B}}_{b,b'=1} \mathcal{E}_{aba'b'} 
 \hat{M}_a \otimes  \hat{N}_b \, \rho \, 
 \hat{M}^\dagger_{a'} \otimes  \hat{N}^\dagger_{b'},
\end{equation}
where we defined the coefficients
$\mathcal{E}_{aba'b'}=\sum_\ell k^{(\ell)}_{ab} k^{(\ell)*}_{a'b'}$, which satisfy
$\mathcal{E}_{abab}=1$ by the trace preserving and 
$\sum_{a,b,a',b'} Z_{ab}\mathcal{E}_{aba'b'}Z^*_{a'b'} \geq 0 $ for all complex numbers 
$Z_{ab}$ by the definition itself.
Next, we show the necessary condition. 
We assume that a map 
$\Phi$ is given as
\begin{equation}
\Phi[\rho]
=
\sum^{d_\text{A}}_{a,a'=1} 
 \sum^{d_\text{B}}_{b,b'=1}
\mathcal{E}_{aba'b'} 
\hat{M}_a \otimes  \hat{N}_b \, \rho \, 
\hat{M}^\dagger_{a'} \otimes  \hat{N}^\dagger_{b'}, 
\label{eq:Phi2}
\end{equation}
where 
$\mathcal{E}_{aba'b'}$  satisfies 
$\mathcal{E}_{abab}=1$ and 
$\sum_{a,b,a',b'} Z_{ab}\mathcal{E}_{aba'b'}Z^*_{a'b'} \geq 0$ for all complex numbers 
$Z_{ab}$.
From the condition 
$\mathcal{E}_{abab}=1$ and the form of the map 
$\Phi$, it is easy to check that 
$\Phi$ is a trace-preserving and population-preserving map.
By the other condition 
$\sum_{a,b,a',b'} Z_{ab}\mathcal{E}_{aba'b'}Z^*_{a'b'} \geq 0 $, we can always find
the complex numbers
$k^{(\ell)}_{ab}$
such that 
$\mathcal{E}_{aba'b'}=\sum_\ell k^{(\ell)}_{ab} k^{(\ell)*}_{a'b'}$. 
Thus, 
\begin{align}
\Phi[\rho]
&
=
\sum^{d_\text{A}}_{a,a'=1} 
 \sum^{d_\text{B}}_{b,b'=1}
\mathcal{E}_{aba'b'} 
\hat{M}_a \otimes  \hat{N}_b \, \rho \, 
\hat{M}^\dagger_{a'} \otimes  \hat{N}^\dagger_{b'} 
\nonumber 
\\
&
=
\sum^{d_\text{A}}_{a,a'=1} 
 \sum^{d_\text{B}}_{b,b'=1}
\sum_\ell k^{(\ell)}_{ab} k^{(\ell)*}_{a'b'}
\hat{M}_a \otimes  \hat{N}_b \, \rho \, 
\hat{M}^\dagger_{a'} \otimes  \hat{N}^\dagger_{b'} 
\nonumber 
\\
&
=
\sum_\ell
\Big(\sum^{d_\text{A}}_{a=1} 
 \sum^{d_\text{B}}_{b=1}
k^{(\ell)}_{ab}
\hat{M}_a \otimes  \hat{N}_b\Big) \, \rho \, 
\Big(\sum^{d_\text{A}}_{a'=1} 
 \sum^{d_\text{B}}_{b'=1} k^{(\ell)}_{a'b'}\hat{M}_{a'} \otimes  \hat{N}_{b'} \Big)^\dagger.
\label{eq:Phi3}
\end{align}
This corresponds to the operator-sum representation. 
Hence, 
$\Phi$ is a population-preserving CPTP map.
\end{proof}
\end{theorem}

Theorem 1 gives the representation of CPTP maps on two particles each in a superposition of multiple paths (two paths when  
$d_\text{A}=d_\text{B}=2$). 
In the next subsection, using Theorem 1, we characterize separable operations on spatially superposed particles. 

\subsection{Characterization of separable operations}
\label{subsec:sepB}

We define a separable operation $\Phi_\text{SEP}$ \cite{Rains1997,Vedral1998} on a bipartite system AB as
\begin{equation}
 \Phi_\text{SEP}[\rho]=
 \sum_\ell \hat{A}_\ell \otimes \hat{B}_\ell \, \rho \, \hat{A}^\dagger_\ell \otimes \hat{B}^\dagger_\ell, 
\end{equation}
where 
$\rho$ is a density operator on the Hilbert space
$\mathcal{H}_\text{A} \otimes \mathcal{H}_\text{B}$, and
the operators 
$\hat{A}_\ell$ and 
$\hat{B}_\ell$ locally act on each Hilbert space and satisfy 
$\sum_\ell \hat{A}^\dagger_\ell \hat{A}_\ell \otimes \hat{B}^\dagger_\ell \hat{B}_\ell =\hat{\mathbb{I}}$.
Since 
$\hat{A}_\ell$ 
and 
$\hat{B}_\ell$ depend on the same parameter
$\ell$ (in general, it may be continuous), the systems A and B are correlated. 
However, quantum entanglement is not generated by separable operations, since one can show that separable operations transform a separable state into an another separable state.
It is known that 
every LOCC operation is a separable operation \cite{Horodecki2009}. 
In terms of LOCC, the parameter 
$\ell$ plays a role of classical information (e.g. measurement outcomes) exchanged through classical channels. 

The crucial theorem to determine whether a given CPTP map is separable was proved in \cite{Cirac2001}.
To explain the theorem, let me introduce the following matrix (operator) 
\begin{equation}
 \mathcal{E}_\Phi =  \Phi \otimes \mathcal{I}_{\text{A}'\text{B}'} 
 \Big[ |\Psi \rangle_{\text{AA}'} \langle \Psi | \otimes |\tilde{\Psi} \rangle_{\text{BB}'} \langle \tilde{\Psi} | 
 \Big], 
 \label{eq:Choi}
\end{equation}
where 
$\Phi$ is a CPTP map for a bipartite system AB with a Hilbert space 
$\mathcal{H}_\text{A} \otimes \mathcal{H}_\text{B}$, and
$\mathcal{I}_{\text{A}'\text{B}'}$ is the identity map for two extra systems 
$\text{A}'$ and 
$\text{B}'$. 
Here, 
$|\Psi \rangle_{\text{AA}'}$ and 
$|\tilde{\Psi} \rangle_{\text{BB}'}$
are the maximally entangled (unnormalized) states as 
\begin{equation}
 |\Psi \rangle_{\text{AA}'}= \sum^{d}_{a=1} |a \rangle_\text{A} |a \rangle_{\text{A}'}, 
 \quad  |\tilde{\Psi} \rangle_{\text{BB}'}= \sum^{d'}_{b=1} |b \rangle_\text{B} |b \rangle_{\text{B}'},
\end{equation}
where 
$d=\dim \mathcal{H}_\text{A}$ and
$d'=\dim \mathcal{H}_\text{B}$.
The properties of the map 
$\Phi$ is translated into those of the matrix
$\mathcal{E}_\Phi$.
In Ref. \cite{Cirac2001}, the matrix 
$\mathcal{E}_\Phi$ was introduced as an extension of the Choi-Jamiolkowski isomorphism \cite{Jamiolkowski1972, Choi1975}, which leads to the following statement. 
\begin{theorem}(the operation separability \cite{Cirac2001})
a CPTP map 
$\Phi$ is a separable operation if and only if the matrix 
$\mathcal{E}_\Phi$ given in Eq.\eqref{eq:Choi} is a separable density matrix with respect to 
$\text{AA}'$ and 
$\text{BB}'$ (up to the normalization constant). 
\end{theorem}
Before applying Theorem 2, it is useful to state the following lemma obtained from Theorem 1. 
\begin{lemma}
Let 
$\Phi$ be the population-preserving CPTP map in Theorem 1. 
The matrix 
$\mathcal{E}_\Phi$ given in Eq.\eqref{eq:Choi} is the square matrix 
$\mathcal{E}$ defined in Theorem 1. 
\begin{proof}
By Theorem 1, a population-preserving CPTP map
$\Phi$ is represented as 
\begin{equation}
\Phi[\rho_\text{in}]
 =\sum^{d_\text{A}}_{a,a'=1} 
 \sum^{d_\text{B}}_{b,b'=1} \mathcal{E}_{aba'b'} 
 \hat{M}_a \otimes  \hat{N}_b \, \rho_\text{in} \, 
 \hat{M}^\dagger_{a'} \otimes  \hat{N}^\dagger_{b'}, 
\label{eq:CPTPPP3}
\end{equation}
with
$\hat{M}_a=|a\,;\text{out}\rangle_\text{A} \langle a\,;\text{in}|$, 
$\hat{N}_b=|b\,;\text{out} \rangle_\text{B} \langle b\,;\text{in}|$, the conditions 
$\mathcal{E}_{abab}=1$ and 
$\sum_{a,b,a',b'} Z_{ab}\mathcal{E}_{aba'b'}Z^*_{a'b'} \geq 0 $ for all complex numbers 
$Z_{ab}$. 
The corresponding Choi matrix is 
\begin{align}
\mathcal{E}_\Phi 
&= 
\Phi \otimes \mathcal{I}_{\text{A}'\text{B}'} 
\Big[ 
|\Psi \rangle_{\text{AA}'} \langle \Psi | \otimes |\tilde{\Psi} \rangle_{\text{BB}'} \langle \tilde{\Psi} | 
\Big]
\nonumber 
\\
&= 
\sum^{d_\text{A}}_{a,a'=1} 
 \sum^{d_\text{B}}_{b,b'=1} \mathcal{E}_{aba'b'} 
\hat{M}_a|\Psi \rangle_{\text{AA}'} \langle \Psi |\hat{M}^\dagger_{a'}
\otimes 
\hat{N}_b|\tilde{\Psi} \rangle_{\text{BB}'} \langle \tilde{\Psi} | \hat{N}^\dagger_{b'}
\nonumber 
\\
&= 
\sum^{d_\text{A}}_{a,a'=1} 
 \sum^{d_\text{B}}_{b,b'=1} \mathcal{E}_{aba'b'} 
|\Psi_a \rangle_{\text{AA}'} \langle \Psi_{a'} |
\otimes 
|\tilde{\Psi}_b \rangle_{\text{BB}'} \langle \tilde{\Psi}_{b'} |, 
\label{eq:EPhi}
\end{align}
where 
$|\Psi_a\rangle_{\text{AA}'}=\hat{M}_a|\Psi \rangle_{\text{AA}'}$ and
$|\tilde{\Psi}_b\rangle_{\text{BB}'}=\hat{N}_b|\tilde{\Psi} \rangle_{\text{BB}'}$. 
The maximally entangled (unnormalized) states 
$|\Psi \rangle_{\text{AA}'}$ and 
$|\tilde{\Psi} \rangle_{\text{BB}'}$ are defined as 
\begin{align}
|\Psi \rangle_{\text{AA}'} 
=\sum^{d_\text{A}}_{a=1} 
|a\,;\text{in} \rangle_\text{A} |a \rangle_{\text{A}'}, \quad 
|\tilde{\Psi} \rangle_{\text{BB}'} =
 \sum^{d_\text{B}}_{b=1} |b\,;\text{in} \rangle_\text{B} |b \rangle_{\text{B}'},
 \label{eq:me}
\end{align}
where 
$|a \rangle_{\text{A}'}$ and 
$|b \rangle_{\text{B}'}$ are the states of extra systems 
$\text{A}'$ and 
$\text{B}'$ satisfying 
${}_{\text{A}'}\langle a'|a \rangle_{\text{A}'}=\delta_{a'a}$ and 
${}_{\text{B}'}\langle b'|b \rangle_{\text{B}'}=\delta_{b'b}$, respectively. 
Since we find the orthonormality
${}_{\text{AA}'} \langle \Psi_{a'}|\Psi_a\rangle_{\text{AA}'}=\delta_{a'a}$ and 
${}_{\text{BB}'} \langle \tilde{\Psi}_{b'}|\tilde{\Psi}_b\rangle_{\text{BB}'}=\delta_{b'b}$, the components of the matrix 
$\mathcal{E}_\Phi$ are
$\mathcal{E}_{aba'b'}$, that is, 
$\mathcal{E}=\mathcal{E}_\Phi$.
\end{proof}
\end{lemma}

Using Theorem 1, Lemma 1 and Theorem 2, we can show the statement on separable operations as follows. 
\begin{theorem}
Let 
$\Phi$ be the population-preserving CPTP map in Theorem 1. 
The map 
$\Phi$ is a separable operation if and only if the matrix 
$\mathcal{E}$ defined in Theorem 1 with the components 
$\mathcal{E}_{aba'b'}$ has the following separable form as
\begin{equation}
\mathcal{E}_{aba'b'}
=\sum_{k} \lambda_k \, (\mathcal{A}_k)_{aa'} \, (\mathcal{B}_k)_{bb'},
\label{eq:sep2}
\end{equation}
where 
$\lambda_k \geq 0$, and 
the matrices 
$\mathcal{A}_k$ and 
$\mathcal{B}_k$ with the components 
$(\mathcal{A}_k)_{aa'}$ and 
$(\mathcal{B}_k)_{bb'}$ have only non-negative eigenvalues. 
\begin{proof}
We assume that a map 
$\Phi$ is CPTP and population-preserving.
As shown in Eq.\eqref{eq:EPhi}, the matrix $\mathcal{E}_\Phi$ associated with the map $\Phi$ is given as 
\begin{equation}
\mathcal{E}_\Phi 
=\sum^{d_\text{A}}_{a,a'=1} 
 \sum^{d_\text{B}}_{b,b'=1} \mathcal{E}_{aba'b'} 
|\Psi_a \rangle_{\text{AA}'} \langle \Psi_{a'} |
\otimes 
|\tilde{\Psi}_b \rangle_{\text{BB}'} \langle \tilde{\Psi}_{b'} |, 
\label{eq:EPhi2}
\end{equation}
where 
$|\Psi_a\rangle_{\text{AA}'}=\hat{M}_a|\Psi \rangle_{\text{AA}'}$ and
$|\tilde{\Psi}_b\rangle_{\text{BB}'}=\hat{N}_b|\tilde{\Psi} \rangle_{\text{BB}'}$ with 
$\hat{M}_a=|a\,;\text{out}\rangle_\text{A} \langle a\,;\text{in}|$, 
$\hat{N}_b=|b\,;\text{out} \rangle_\text{B} \langle b\,;\text{in}|$ and the maximally entangled (unnormalized) states 
$|\Psi \rangle_{\text{AA}'}$ and 
$|\tilde{\Psi} \rangle_{\text{BB}'}$ defined in \eqref{eq:me}. 
If 
$\Phi$ is separable, then, by Theorem 2, 
$\mathcal{E}_\Phi$ has the separable form 
\begin{equation}
\mathcal{E}_\Phi=\sum_{k} q_k \tilde{\rho}_k \otimes \tilde{\sigma}_k,
\end{equation}
where 
$q_k \geq 0$, and 
$\tilde{\rho}_k$ and 
$\tilde{\sigma}_k$ are density operators on 
$\mathcal{H}_\text{A} \otimes \mathcal{H}_{\text{A}'}$ and 
$\mathcal{H}_\text{B} \otimes \mathcal{H}_{\text{B}'}$, respectively. 
By the orthonormality
${}_{\text{AA}'} \langle \Psi_{a'}|\Psi_a\rangle_{\text{AA}'}=\delta_{a'a}$ and 
${}_{\text{BB}'} \langle \tilde{\Psi}_{b'}|\tilde{\Psi}_b\rangle_{\text{BB}'}=\delta_{b'b}$ and 
using Eq.\eqref{eq:EPhi2}, we have
\begin{equation}
\mathcal{E}_{aba'b'}
=
{}_{\text{AA}'}\langle \Psi_a| {}_{\text{BB}'} \langle \tilde{\Psi}_b |
 \mathcal{E}_\Phi 
 |\Psi_{a'} \rangle_{\text{AA}'} 
 |\tilde{\Psi}_{b'} \rangle_{\text{BB}'}
 = \sum_{k} q_k (\tilde{\rho}_k)_{aa'}(\tilde{\sigma}_k)_{bb'},
\end{equation}
where 
$(\tilde{\rho}_k)_{aa'}={}_{\text{AA}'}\langle \Psi_a|\tilde{\rho}_k  |\Psi_{a'}\rangle_{\text{AA}'}$ and 
$(\tilde{\sigma}_k)_{bb'}={}_{\text{BB}'}\langle \tilde{\Psi}_b|\tilde{\sigma}_k  |\tilde{\Psi}_{b'}\rangle_{\text{BB}'}$.
The form of
$\mathcal{E}_{aba'b'}$ is nothing but the separable form. 
Conversely, 
if 
$\mathcal{E}_{aba'b'}$ has the following separable form
\begin{equation}
\mathcal{E}_{aba'b'}
=\sum_{k} \lambda_k \, (\mathcal{A}_k)_{aa'} \, (\mathcal{B}_k)_{bb'},
\label{eq:sep2p}
\end{equation}
where 
$\lambda_k \geq 0$, and 
the matrices 
$\mathcal{A}_k$ and 
$\mathcal{B}_k$ with the components 
$(\mathcal{A}_k)_{aa'}$ and 
$(\mathcal{B}_k)_{bb'}$ have only non-negative eigenvalues, then the matrix 
$\mathcal{E}_\Phi$ given in \eqref{eq:EPhi2} is a separable density matrix with respect to 
$\text{AA}'$ and
$\text{BB}'$ (up to normalization constant). 
By Theorem 2, the population-preserving CPTP map 
$\Phi$ is separable. 
\end{proof}
\end{theorem}

In particular, if the dimension of the Hilbert space in Theorem 3 is six or less
(a two-qubit or qubit-qutrit system), the following corollary is obtained. 
\begin{corollary}
In Theorem 3, we further assume that the dimension of the total Hilbert space satisfies
$d_\text{A} d_\text{B} \leq 6$.
Then, a population-preserving CPTP map 
$\Phi$ is separable if and only if the partial transposition
$\mathcal{E}^{\text{T}_\text{A}}$ with the elements 
$(\mathcal{E}^{\text{T}_\text{A}})_{aba'b'}=\mathcal{E}_{a'bab'}$ 
has only non-negative eigenvalues. 
\end{corollary}
This corollary follows by the fact that, when
$d_\text{A}d_\text{B} \leq6$, the PPT criterion for the matrix 
$\mathcal{E}_\Phi (=\mathcal{E})$ is the necessary and sufficient condition on the separability of  
$\mathcal{E}_\Phi$.
Corollary 1 gives us the following criterion on inseparable operations. 
We have a population-preserving CPTP map 
$\Phi$ on two particles. 
We assume that one of the particles is superposed in two paths and the other is in two or three paths. 
The partial transposition 
$\mathcal{E}^{\text{T}_\text{A}}$ of the square matrix 
$\mathcal{E}$ has a negative eigenvalue if and only if the map 
$\Phi$ is inseparable. 

\begin{figure}[H]
  \centering
  \includegraphics[width=0.80\linewidth]{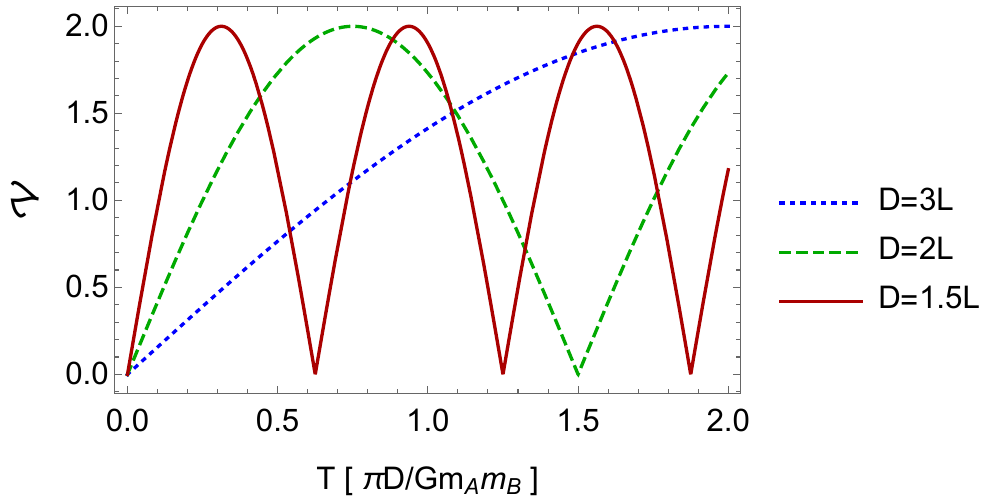}
  \caption{The behavior of the measure $\mathcal{V}$ to test the inseparable operation due to the gravitational interaction between two massive particles.
  The vertical axis represents the dimensionless quantity 
  $\mathcal{V}$. 
  The horizontal axis describes the time 
  $T$ in the unit 
  $\pi D /Gm_\text{A}m_\text{B}$.
  }
  \label{fig:chngtvty}
\end{figure}

To demonstrate how to use the criterion, let us focus on the dynamics of superposed massive particles considered in Sec.\ref{sec:dynamics}. 
The evolved state given by Eq.\eqref{eq:Sol} is rewritten as 
\begin{equation}
|\Psi_\text{out} \rangle =
\sum_{a,b=\text{L},\text{R}}e^{i\Phi_{ab}}\, \psi_a |a\,;\text{out}\rangle_\text{A} \, \phi_b |b\,;\text{out} \rangle_\text{B}
=\sum_{a,b=\text{L},\text{R}}e^{i\Phi_{ab}}\, \hat{M}_a \otimes \hat{N}_b | \Psi_\text{in} \rangle,
\label{eq:Sol2}
\end{equation}
or equivalently, 
\begin{equation}
|\Psi_\text{out} \rangle \langle \Psi_\text{out}| 
=\sum_{a,a'=\text{L},\text{R}}
\sum_{b,b'=\text{L},\text{R}}
e^{i(\Phi_{ab}-\Phi_{a'b'})}\, \hat{M}_a \otimes \hat{N}_b 
|\Psi_\text{in} \rangle \langle \Psi_\text{in}| \hat{M}^\dagger_{a'} \otimes \hat{N}^\dagger_{b'},
\label{eq:Sol3}
\end{equation}
where 
$|\Psi_\text{in}\rangle$ is the initial state \eqref{eq:Psii}, 
$\hat{M}_a=|a\,;\text{out}\rangle_\text{A} \langle a\,;\text{in}|$ and 
$\hat{N}_b=|b\,;\text{out} \rangle_\text{B} \langle b\,;\text{in}|$. 
Then, we find the matrix 
$\mathcal{E}$ with the elements $\mathcal{E}_{aba'b'}=e^{i(\Phi_{ab} -\Phi_{a'b'})}$ as
\begin{equation}
\mathcal{E}=
\begin{bmatrix}
1 & e^{i(\Phi_\text{LL} -\Phi_\text{LR})} & e^{i(\Phi_\text{LL} -\Phi_\text{RL})} & e^{i(\Phi_\text{LL} -\Phi_\text{RR})} \\
e^{i(\Phi_\text{LR} -\Phi_\text{LL})} & 1 &  e^{i(\Phi_\text{LR} -\Phi_\text{RL})} & e^{i(\Phi_\text{LR} -\Phi_\text{RR})} \\
e^{i(\Phi_\text{RL} -\Phi_\text{LL})} & e^{i(\Phi_\text{RL} -\Phi_\text{LR})} & 1 & e^{i(\Phi_\text{RL} -\Phi_\text{RR})} \\
e^{i(\Phi_\text{RR} -\Phi_\text{LL})} & e^{i(\Phi_\text{RR} -\Phi_\text{LR})} &  e^{i(\Phi_\text{RR} -\Phi_\text{RL})} & 1 \\
\end{bmatrix}, 
\label{eq:E}
\end{equation}
where the 4 rows and 4 columns labeled by 
$ab=\text{LL},\text{LR},\text{RL},\text{RR}$ and 
$a'b'=\text{LL},\text{LR},\text{RL},\text{RR}$, respectively.
The partial transposition 
$\mathcal{E}^{\text{T}_\text{A}}$ with the components  
$(\mathcal{E}^{\text{T}_\text{A}})_{aba'b'}=\mathcal{E}_{a'bab'}$
is 
\begin{equation}
\mathcal{E}^{\text{T}_\text{A}}=
\begin{bmatrix}
1 & e^{i(\Phi_\text{LL} -\Phi_\text{LR})} & e^{i(\Phi_\text{RL} -\Phi_\text{LL})} & e^{i(\Phi_\text{RL} -\Phi_\text{LR})} \\
e^{i(\Phi_\text{LR} -\Phi_\text{LL})} & 1 &  e^{i(\Phi_\text{RR} -\Phi_\text{LL})} & e^{i(\Phi_\text{RR} -\Phi_\text{LR})} \\
e^{i(\Phi_\text{LL} -\Phi_\text{RL})} & e^{i(\Phi_\text{LL} -\Phi_\text{RR})} & 1 & e^{i(\Phi_\text{RL} -\Phi_\text{RR})} \\
e^{i(\Phi_\text{LR} -\Phi_\text{RL})} & e^{i(\Phi_\text{LR} -\Phi_\text{RR})} &  e^{i(\Phi_\text{RR} -\Phi_\text{RL})} & 1 \\
\end{bmatrix}.
\label{eq:ETA}
\end{equation}
In the same manner as the definition of the negativity, the measure to test the inseparable operation is introduced as 
\begin{equation}
\mathcal{V}=\sum_{\nu_i <0} |\nu_i|,
\label{eq:V}
\end{equation}
where 
$\nu_i$ are the eigenvalues of 
$\mathcal{E}^{\text{T}_\text{A}}$. 
By a straightforward computation, the measure 
$\mathcal{V}$ is obtained as 
\begin{equation}
\mathcal{V}
=2 \Big|\sin \Big(\frac{\Phi_\text{LR}+\Phi_\text{RL}-\Phi_\text{LL}-\Phi_\text{RR}}{2} \Big) \Big|
=2\Big|
 \sin
\Big[
\frac{Gm_\text{A}m_\text{B}T}{2} \Big( \frac{1}{D+L}+\frac{1}{D-L}-\frac{2}{D} \Big)
\Big]\Big|,
\label{eq:expV}
\end{equation}
where the formula \eqref{eq:PhiPQ2} of 
$\Phi_{ab}$ was substituted. 
Fig.\ref{fig:chngtvty} presents the measure as a function of 
$Gm_\text{A}m_\text{B}T/\pi D$ for a fixed 
$D$. 
This measure 
$\mathcal{V}$ is proportional to the negativity 
$\mathcal{N}$.
The inseparability of the operation due to the gravitational interaction is consistent with the entanglement generation due to it. 

\subsection{Equivalence between inseparable and entangling operations}
\label{subsec:eqC}

Let us introduce a larger class of evolutions called non-entangling operations \cite{Harrow2003}. 
A CPTP map 
$\Phi_\text{NE}$ is called a non-entangling operation if 
$\Phi_\text{NE}[\rho]$ is a separable state for every separable state 
$\rho$. 
It is obvious that separable operations are also non-entangling. 
The converse statement is not true because of counterexamples. 
In general, separable 
operations are not equivalent to non-entangling operations.

For example, the SWAP operation
$\hat{U} \, \rho \otimes \sigma \hat{U}^\dagger=\sigma \otimes \rho$ 
is not separable but non-entangling. 
Indeed, we have 
\begin{equation}
\Phi_U \Big[\sum_{k} p_k \rho_k \otimes \sigma_k \Big]
=\hat{U} \sum_{k} p_k \rho_k \otimes \sigma_k \hat{U}^\dagger
=\sum_{k} p_k \hat{U}  \rho_k \otimes \sigma_k \hat{U}^\dagger
=\sum_{k} p_k \sigma_k \otimes \rho_k,
\label{eq:nonent}
\end{equation}
hence the SWAP operation 
$\Phi_U$ is non-entangling. 
On the other hand, using \eqref{eq:Choi} for $\Phi_U$, we find the matrix
\begin{align}
 \mathcal{E}_U
&= 
\Phi_U \otimes \mathcal{I}_{\text{A}'\text{B}'} 
 \Big[ |\Psi \rangle_{\text{AA}'} \langle \Psi | \otimes |\Psi \rangle_{\text{BB}'} \langle \Psi | 
 \Big]
\nonumber 
\\
&= 
\sum^d_{i,i',j,j'=1} 
\Phi_U \otimes \mathcal{I}_{\text{A}'\text{B}'} 
 \Big[ 
|i \rangle_\text{A}  \langle i' | \otimes |i \rangle_{\text{A}'} \langle i' | 
 \otimes 
|j \rangle_\text{B}  \langle j' | \otimes |j \rangle_{\text{B}'} \langle j' |
\Big]
\nonumber 
\\
&= 
\sum^d_{i,i',j,j'=1}  
|j \rangle_\text{A}  \langle j' | \otimes |i \rangle_{\text{A}'} \langle i' | 
 \otimes 
|i \rangle_\text{B}  \langle i' | \otimes |j \rangle_{\text{B}'} \langle j' |
\nonumber 
\\
&= 
|\Psi \rangle_{\text{AB}'} \langle \Psi | \otimes |\Psi \rangle_{\text{A}'\text{B}} \langle \Psi|
 \label{eq:insep}
\end{align}
where 
$|\Psi \rangle
=\sum^{d}_{\text{i}=1} 
|i\rangle |i \rangle$. 
The matrix 
$\mathcal{E}_U$ is entangled with respect to 
$\text{AA}'$ and
$\text{BB}'$. 
By Theorem 2, the SWAP operation 
$\Phi_U$ is not separable . 

For realizing inseparable 
or non-LOCC 
operations on two quantum systems, a quantum channel (a channel to transfer quantum information) between the systems may be required.
The SWAP operation is an example of a quantum channel without entanglement generation.  
On the other hand, we find the following statement on inseparable and entangling operations. 
\begin{theorem}
Let 
$\Phi$ be the population-preserving CPTP map in Theorem 1. 
The map 
$\Phi$ is inseparable if and only if  the map 
$\Phi$ is entangling. 
\begin{proof}
Since every entangling operation is inseparable, we consider the converse statement. 
Let a population-preserving CPTP map
$\Phi$ be an inseparable operation. 
By Theorem 3, 
the matrix 
$\mathcal{E}$ with the elements $\mathcal{E}_{aba'b'}$
has no separable form.  
Assuming 
$\rho_\text{in}=|\psi_\text{in} \rangle_\text{A} \langle \psi_\text{in}| \otimes |\phi_\text{in} \rangle_\text{B} \langle \phi_\text{in}|$ and 
using Theorem 1, we find 
\begin{align}
\Phi[\rho_\text{in}]
&=\Phi[|\psi_\text{in} \rangle_\text{A} \langle \psi_\text{in}| \otimes |\phi_\text{in} \rangle_\text{B} \langle \phi_\text{in} |]
\nonumber 
\\
&=\sum^{d_\text{A}}_{a,a'=1} 
 \sum^{d_\text{B}}_{b,b'=1}
\mathcal{E}_{aba'b'}
\hat{M}_a \otimes \hat{N}_b 
\, |\psi_\text{in} \rangle_\text{A} \langle \psi_\text{in}| \otimes |\phi_\text{in} \rangle_\text{B} \langle \phi_\text{in}| \,
\hat{M}^\dagger_{a'} \otimes \hat{N}^\dagger_{b'}
\nonumber 
\\
&=\sum^{d_\text{A}}_{a,a'=1} 
 \sum^{d_\text{B}}_{b,b'=1}
\mathcal{E}_{aba'b'} 
\psi^*_a \psi_{a'}
\phi^*_b \phi_{b'}
| a\,;\text{out} \rangle_\text{A} \langle a';\text{out} | \otimes | b\,;\text{out} \rangle_\text{B} \langle b';\text{out} |, 
\label{eq:Phirho}
\end{align}
where 
$\psi_a={}_\text{A} \langle a\,;\text{in}|\psi_\text{in} \rangle_\text{A}$ and 
$\phi_b={}_\text{B} \langle b\,;\text{in}|\phi_\text{in} \rangle_\text{B}$. 
We can always choose 
$|\psi_\text{in} \rangle_\text{A}$ and 
$|\phi_\text{in} \rangle_\text{B}$ so that  
$\psi_a =1/\sqrt{d_\text{A}} $
and 
$\phi_b=1/\sqrt{d_\text{B}}$, and then the elements of the density matrix 
$\Phi[\rho_\text{in}]$ are given by
$\mathcal{E}_{aba'b'}/d_\text{A}d_\text{B}$. 
Since 
$\mathcal{E}_{aba'b'}$ is not separable, the density matrix 
$\Phi[\rho_\text{in}]$ describes an entangled state, and hence 
$\Phi$ is an entangling operation. 
\end{proof}
\end{theorem}
Theorem 4 implies that inseparable operations on two particles in a superposition of paths are entangling operations and that the associated quantum channel can create entanglement.

The theorems obtained so far are independent of the details of the interaction between two particles. 
Applying the theorems, we find the following implication for quantum interaction induced by gravity. 
Whenever a gravitational interaction is described by an inseparable operation on path-superposed masses, the gravitational field mediating the interaction creates entanglement. 
The quantum channel model of gravity described by an inseparable operation without entanglement generation is incompatible with the possible evolution of the masses.

\section{Conclusion}
\label{sec:Conclusion}

To understand the interaction via the gravitational field induced by superposed particles, we investigated when entanglement occurs between two particles in a path superposition. 
The present approach based on quantum information theory allowed us to analyze the evolution of the particles without specifying a model of the interaction between them.
The representation theorem on the evolution given by a population-preserving CPTP map identifies separable operations on the superposed particles. 
As a byproduct, the simple procedure to test the separability criterion on the evolution was obtained. 
We further found that inseparable operations on the particles are nothing but non-entangling operations. 
This means that the entanglement between the particles occurs only when the quantum interaction between them induces an inseparable operation. 
If the interaction is mediated by a gravitational field, we have the robust statement: there are no quantum models of gravitational interaction given by an inseparable operation which does not generate entanglement between two masses in a path superposition. 

In an experimental viewpoint, the present setting of quantum systems in this paper have respected matter-wave interferometers. 
The studies on other proposed systems, for example, levitated nanoparticles and macroscopic mechanical oscillators will be left as the future works.
It is interesting to see what kind of quantum channels the gravitational field forms in such  physical systems.


\begin{acknowledgments}
We thank S. Kanno, S. Kukita, Y. Kuramochi and K. Yamamoto for useful discussions and comments related to this paper. 
\end{acknowledgments}



\end{document}